\documentclass[conference]{IEEEtran}
\usepackage{amsfonts}
\usepackage{amsmath}
\usepackage{graphicx}
\usepackage{url}
\usepackage{eepic, epsfig, amsmath, amssymb, latexsym, setspace, subfig}
\usepackage{rotating}
\usepackage{amsfonts}
\usepackage{amsmath}
\usepackage{graphicx}
\usepackage{nicefrac}
\usepackage{lipsum,multicol}
\usepackage{authblk}

\newtheorem{algorithm}{Algorithm}

\numberwithin{equation}{section}

\newcommand{\qed}{\rule{7pt}{7pt}}

\newcommand{\x}{\mathbf{x}}
\newcommand{\y}{\mathbf{y}}
\newcommand{\z}{\mathbf{z}}
\newcommand{\w}{\mathbf{w}}

\newcommand{\A}{{\bf A}}
\def\x{{\bf x}}
\def\y{{\bf y}}
\def\w{{\bf w}}

\def\z{{\bf z}}

\newcommand{\beq}{\begin{equation}}
\newcommand{\eeq}{\end{equation}}
\newcommand{\bea}{\begin{eqnarray}}
\newcommand{\eea}{\end{eqnarray}}

\newcommand{\Prob}{\ensuremath{\mathbb{P}}}
\def\qed{\quad \vrule height6.5pt width6pt depth0pt} 

\def\qed{\quad \vrule height6.5pt width6pt depth0pt} 

\newtheorem{thm}{Theorem}

\IEEEoverridecommandlockouts
\begin{document}
%

\title{Universally Elevating the Phase Transition Performance of Compressed Sensing: Non-Isometric Matrices are Not Necessarily Bad Matrices}
\author[1]{Weiyu Xu}
\author[2]{Myung Cho}

\affil[1]{Dept. of ECE, University of Iowa\\
Email: weiyu-xu@uiowa.edu}
\affil[2]{Dept. of ECE,
University of Iowa\\
Email: myung-cho@uiowa.edu}
\maketitle
\begin{abstract}
In compressed sensing problems, $\ell_1$ minimization or Basis Pursuit was known to have the best provable phase transition performance of recoverable sparsity among polynomial-time algorithms. It is of great theoretical and practical interest to find alternative polynomial-time algorithms which perform better than $\ell_1$ minimization. \cite{Icassp reweighted l_1}, \cite{Isit reweighted l_1}, \cite{XuScaingLaw} and \cite{iterativereweightedjournal} have shown that a two-stage re-weighted $\ell_1$ minimization algorithm can boost the phase transition performance for signals whose nonzero elements follow an amplitude probability density function (pdf) $f(\cdot)$ whose $t$-th derivative $f^{t}(0) \neq 0$ for some integer $t \geq 0$. However, for signals whose nonzero elements are strictly suspended from zero in distribution (for example, constant-modulus, only taking values `$+d$' or `$-d$' for some nonzero real number $d$), no polynomial-time signal recovery algorithms were known to provide better phase transition performance than plain $\ell_1$ minimization, especially for dense sensing matrices. In this paper, we show that a polynomial-time algorithm can universally elevate the phase-transition performance of compressed sensing, compared with $\ell_1$ minimization, even for signals with constant-modulus nonzero elements. Contrary to conventional wisdoms that compressed sensing matrices are desired to be isometric, we show that non-isometric matrices are not necessarily bad sensing matrices. In this paper, we also provide a framework for recovering sparse signals when sensing matrices are not isometric.

\end{abstract}

\IEEEpeerreviewmaketitle

\section{Introduction}
\label{sec:Intro}
Compressed sensing addresses the problem of recovering sparse signals
from under-determined systems of linear equations  \cite{CT, D CS}. In
particular, if $\x$ is an $n\times1$ real-numbered vector that is known to have at
most $k$ nonzero elements where $k<n$, and $\A$ is an $m\times n $
measurement matrix with $k<m<n$, then for appropriate values of $k$,
$m$ and $n$, it is possible to efficiently recover $\x$ from $\y=\A\x$ \cite{D
  CS,DT,CT}. In this paper, we only consider the asymptotically linear case where $\frac{m}{n} \rightarrow \delta$ for a certain constant $\delta>0$. For simplicity of presentation, we only consider the case that the observation $\y$ is noiseless.  One of the most important recovery
algorithm is $\ell_1$ minimization which can be formulated  as
follows:
\beq
\label{eq:l1 min}
\min_{\A\z=\A\x}\|\z\|_1.
\eeq

Precise phase transitions of signal
recovery using $\ell_1$ minimization were established by Donoho and Tanner in
\cite{DT}\cite{D}\cite{Donoho positive}, by using tools from convex geometry \cite{Santalo1952}\cite{McMullen1975}\cite{Grunbaumbook}. In particular, it was shown, in \cite{DT}\cite{D}\cite{Donoho positive},  that if the measurement matrix have i.i.d. Gaussian elements, for a given ratio
of $\delta = \frac{m}{n}$, $\ell_1$ minimization can successfully
recover {\em every} $k$-sparse signal, provided that $\mu = \frac{k}{n}$ is
smaller that a certain threshold. This successful recovery property holds true with high
probability as $n \rightarrow \infty$.

 This threshold guarantees the recovery of {\em all}
sufficiently sparse signals and is therefore referred to as a ``strong''
threshold. It does not depend on the actual distribution of
the nonzero entries of the sparse signal and thus is a universal
result. 

Another notion introduced and computed in \cite{DT,D} is that of a
{\em weak} threshold $\mu_{W}(\delta)$ under which signal recovery is guaranteed for {\em
  almost all} support sets and {\em almost all} sign patterns of the
sparse signal, with high probability as $n\rightarrow\infty$. The weak
threshold is the phase transition threshold that can be observed in simulations of $\ell_1$
minimization, and allows for signal recovery beyond the strong threshold. It
is also universal in the sense that it is independent of the amplitude of the nonzero elements of sparse signals take.

While a large number of decoding algorithms have been introduced since the breakthrough works \cite{D
  CS,DT,CT}, there were no polynomial-time algorithms which provably provide better phase transition performance than $\mu_{W}(\delta)$ provided by $\ell_1$ minimization, especially for dense sensing matrices. It is worth noting that different variations of reweighted $\ell_1$
algorithms have been recently introduced in the literature \cite{CWB07,Needell, CharYin2008, WY2010sparse, Icassp reweighted l_1,Isit reweighted l_1},  and have shown empirical performance improvements over plain $\ell_1$ minimization for special classes of sparse signals. In \cite{CWB07}, a iterative reweighted $\ell_1$ minimization algorithm was proposed,  and is shown to empirically outperform $\ell_1$
minimization for sparse signals with non-flat
distributions for nonzero elements. However, for signals whose nonzero elements are constant-modulus, namely only taking values `$+d$' or `$-d$' for some nonzero real number $d$, the empirical recoverable sparsity in \cite{CWB07} is almost identical to plain $\ell_1$ minimization. No theoretical results were obtained showing better phase transition performance guarantee than $\ell_1$ minimization in \cite{CWB07}. Similar performance improvements, together with similar limitations for sparse signals with constant-modulus nonzero elements,  were empirically observed for several other variations of iterative reweighted algorithms \cite{CharYin2008, WY2010sparse, Icassp reweighted l_1,Isit reweighted l_1}. These algorithms fail to empirically improve the phase transition performance for signals with constant-modulus nonzero elements, arguably because it is very hard to extract meaningful support information for signals with constant-modulus nonzero elements. In \cite{Needell}, approximately sparse signals have been considered, where perfect recovery is not possible. However, it has been shown that the recovery noise can be reduced using an iterative scheme.

 On the theoretical side, there are results showing that, for certain types of sparse signals, variations of iterative reweighted $\ell_1$ minimization algorithms indeed provide better phase transition performance than plain $\ell_1$ minimization. In particular, in  \cite{Icassp reweighted l_1}, it was shown that a two-stage iterative reweighted $\ell_1$ minimization improves the phase transition performance for a restrictive class of sparse signals whose nonzero elements have dramatically different amplitudes. \cite{Isit reweighted l_1} showed that a two-stage iterative reweighted $\ell_1$ minimization algorithm improves the phase transition performance for sparse signals with Gaussian distributed nonzero elements.
Using a scaling law result for the stability of $\ell_1$ minimization, and the Grassmann angle framework \cite{isitweighted}\cite{weightedJournal} for the weighted $\ell_1$ minimization, \cite{XuScaingLaw} and \cite{iterativereweightedjournal} prove that an iterative reweighted $\ell_1$ algorithm indeed has better phase transition performance for a wide class of sparse signals, including sparse Gaussian signals. The key to these results is that, for these signals, $\ell_1$ minimization has an \emph{approximate support recovery} property \cite{Isit reweighted l_1} which can be exploited by a reweighted $\ell_1$ algorithm, to obtain a provably superior phase transition performance.

More specifically, \cite{XuScaingLaw} and \cite{iterativereweightedjournal} have shown that,  if the nonzero elements over the signal support follow a probability density function (pdf) $f(\cdot)$ whose $t$-th derivative $f^{t}(0) \neq 0$ for some $t \geq 0$, then a certain iterative reweighted $\ell_1$ minimization algorithm can be analytically shown to lift the phase transition thresholds (weak thresholds) of plain $\ell_1$ minimization algorithm through using the scaling law for the sparse recovery stability. In fact, \cite{XuScaingLaw} and \cite{iterativereweightedjournal} extended the results \cite{Isit reweighted l_1} of phase transition improvements for sparse vectors with Gaussian nonzero elements, whose amplitude pdf is nonzero at the origin (namely the pdf's $0$-th derivative is nonzero).  However, \cite{XuScaingLaw} and \cite{iterativereweightedjournal} failed to show phase transition improvement for sparse signals with constant-modulus nonzero elements. Again, this is because the authors were not able to establish approximate support recovery property for sparse signals with constant-modulus nonzero elements. In fact, for sparse signals with nonzero constant-modulus elements, $\ell_1$ minimization is unstable \emph{\emph{as soon as}} the sparsity surpasses the weak threshold, and it is thus very hard to extract support information from the decoding results of plain $\ell_1$ minimization \cite{noisesensitivy}\cite{dynamicL1}\cite{messagepassing}.

In this paper, we do not assume that the decoder has any prior information about the signal support or about the probability density function of the nonzero elements in the sparse signal. When this prior information is available to the decoder, weighted $\ell_1$ minimization \cite{weightedJournal} or message passing algorithms \cite{Spatialcoupling} can improve the phase transition of plain $\ell_1$ minimization.

Naturally, it is of great theoretical and practical interest to find alternative polynomial-time algorithms which perform better than $\ell_1$ minimization, in the absence of any prior information about the sparse signal. Please also see \cite{StojnicQuestion} for discussions on working towards a better compressed sensing. Even though \cite{Icassp reweighted l_1}, \cite{Isit reweighted l_1}, \cite{XuScaingLaw} and \cite{iterativereweightedjournal} have shown that a two-stage re-weighted $\ell_1$ minimization algorithm can boost the phase transition performance for signals whose nonzero elements follow an amplitude probability density function (pdf) $f(\cdot)$ whose $t$-th derivative $f^{t}(0) \neq 0$ for some integer $t \geq 0$, these results are not universal over all the possible probability distributions for nonzero elements. As discussed, the main difficulties are from sparse signals whose nonzero elements are strictly suspended from zero in distribution (for example, constant-modulus, only taking values `$+d$' or `$-d$' for some nonzero real number $d$) \cite{dynamicL1}\cite{messagepassing}\cite{noisesensitivy}.

In this paper, we show that a polynomial-time algorithm can universally elevate the phase-transition performance of compressed sensing, compared with $\ell_1$ minimization, even for sparse signals with constant-modulus nonzero elements. Our ideas is to use non-isometric sensing matrices, and to design modified $\ell_1$ minimization algorithms tailored to these non-isometric sensing matrices. Our theoretical analysis is based on the scaling law for the stability of $\ell_1$ minimization.

Contrary to conventional wisdoms that compressed sensing matrices are desired to be isometric, we show that non-isometric matrices are not necessarily bad sensing matrices. In this paper, we also provide a framework for recovering sparse signals when sensing matrices are not isometric.

This paper is organized as follows. In Section \ref{sec:basic} and \ref{sec:model}, we introduce the basic concepts and system model. In Section \ref{sec:scaling}, we summarize the scaling law \cite{XuScaingLaw} for recovery stability in compressed sensing. In Section \ref{sec:matrices}, we introduce the non-isometric sensing matrices. In Section \ref{sec:Algorithm}, we introduce our new sparse recovery algorithm, and state the main results. In Sections \ref{sec:robustness} and \ref{sec:perfect recovery}, we outline the key steps of our proof. In Section \ref{sec:simulation}, simulation results are given to demonstrate improved phase transition performance, brought by non-isometric sensing matrices and new signal recovery algorithms.
\section{Basic Definitions}
\label{sec:basic}
 A sparse signal with exactly $k$ nonzero entries is called
 $k$-sparse. For a vector $\x$, $\|\x\|_1$ denotes the $\ell_1$
 norm. The support of $\x$,  denoted by $supp(\x)$, is the support
 set of its nonzero coordinates. For a vector $\x$ that is not exactly
 $k$-sparse, we define the $k$-support of $\x$ to be the index set of
 the largest $k$ entries of $\x$ in amplitude, and denote it by
 $supp_k(\x)$. For a subset $K$ of the entries of $\x$, $\x_K$ means
 the vector formed by those entries of $\x$ indexed in $K$. 

\section{Signal Model and Problem Description}
\label{sec:model}
We consider sparse random signals with i.i.d. nonzero
entries. In other words we assume that the unknown sparse signal is an
$n\times 1$ vector $\x$ with exactly $k$ nonzero entries, where each
nonzero entry is independently sampled from a well defined distribution. The measurement matrix $\A$ is an $m\times n$
matrix with a compression ratio $\delta
= \frac{m}{n}$. Compressed sensing theory guarantees that if
$\mu=\frac{k}{n}$ is smaller than a certain threshold, then every
$k$-sparse signal can be recovered using $\ell_1$ minimization. The
relationship between $\delta$ and the maximum threshold of $\mu$ for
which such a guarantee exists is called the \emph{strong sparsity
  threshold} \cite{D}, and is denoted by $\mu_{S}(\delta)$. A more practical
performance guarantee is the so-called \emph{weak sparsity threshold},
denoted by $\mu_{W}(\delta)$, and has the following
interpretation. For a fixed value of $\delta = \frac{m}{n}$ and
i.i.d. Gaussian matrix $\A$ of size $m\times n$,  a random $k$-sparse
vector $\x$ of size $n\times 1$ with a randomly chosen support set and
a random sign pattern can be recovered from $\A\x$ using $\ell_1$
minimization with high probability, if $\frac{k}{n}<\mu_{W}(\delta)$. Other types of recovery thresholds can be
obtained by imposing more or fewer restrictions. For example, strong
and weak thresholds for nonnegative signals have been evaluated in
\cite{Donoho positive}.

We assume that the support size of $\x$, namely $k$, is slightly
larger than the weak threshold of $\ell_1$ minimization. In other
words,  $k = (1+\epsilon_0)\mu_{W}(\delta)$ for some
$\epsilon_0>0$. This means that if we use $\ell_1$ minimization, a
randomly chosen $\mu_{W}(\delta)n$-sparse signal will be recovered
perfectly with very high probability, whereas a randomly selected
$k$-sparse signal will not. We would like to show that for a strictly
positive $\epsilon_0$, the new $\ell_1$ algorithm of
Section \ref{sec:Algorithm} can indeed recover a randomly selected
$k$-sparse signal with high probability, which means that it has an
improved weak threshold.

\section{The Scaling Law for the Compressed Sensing Stability}
\label{sec:scaling}
To prove our sensing matrices and signal recovery algorithms provide better phase transition performance, we need the stability result of compressed sensing when signal sparsity is bigger than the weak threshold. In this section, we will recall from \cite{precisestability, XuScaingLaw} the scaling law of the $\ell_1$ recovery stability as a function of signal sparsity.

When the sparsity of the signal $\x$ is larger than the weak threshold $\mu_{W}(\delta)n$, a common stability result for the $\ell_1$ minimization is that, for a set $K \subseteq \{1,2, ..., n\}$ with cardinality $|K|$ small enough for $A$ to satisfy the restrict isometry condition \cite{CT} or the null space robustness property \cite{devore2,XuHassibiAllerton08}, the decoding error is bounded by,
\begin{equation}
\|\x-\hat{\x}\|_1 \leq D \|\x_{\overline{K}}\|_1,
\label{eq:generalstability}
\end{equation}
where $\hat{\x}$ is any minimizer to $\ell_1$ minimization, $D$ is a constant, $\overline{K}$ is the complement of the set $K$ and $\x_{\overline{K}}$ is the part of $\x$ over the set $\overline{K}$.

To date, known bounds on $|K|/n$, for the restricted isometry condition to hold with overwhelming probability, are small compared with the weak threshold $\mu_{W}(\delta)$ \cite{CT}. \cite{isitrobust} \cite{XuHassibiAllerton08} \cite{precisestability} and \cite{XuScaingLaw} used the Grassmann angle approach to characterize sharp bounds on the stability of $\ell_1$ minimization and showed that, for an arbitrarily small $\epsilon_{0}$, as long as $|K|/n=(1-\epsilon_{0})\mu_{W}(\delta)n$, with overwhelming probability as $n \rightarrow \infty$, (\ref{eq:generalstability}) holds for some constant $D$ ($D$ of course depends on $|K|/n$). In particular, \cite{XuScaingLaw} and \cite{iterativereweightedjournal} gave a \emph{closed-form} characterization for this tradeoff between $C$ (related to $D$), as in the following Theorem \ref{thm:Cstability}, and the sparsity ratio $|K|/n$. This tradeoff is termed as the scaling law for compressive sensing recovery stability, and stated in Theorem \ref{thm:scalinglaw}. First, we first see how recovery stability is related to $C$.

\begin{thm}
\label{thm:Cstability}
Let $A$ be a general $m\times n$ measurement matrix, $\x$ be an $n$-element
vector and $\y=A\x$. Denote $K$ as a subset of $\{1,2,\dots,n\}$ such
that its cardinality $|K|=k$ and further denote $\overline{K}=\{1,2,\dots,n\}\setminus K$. Let $\w$
denote an $n \times 1$ vector. Let $C>1$ be a fixed number.

Given a specific set $K$ and suppose that the part of $\x$ on $K$, namely $\x_{K}$ is fixed.
No matter what $ \x_{\overline{K}}$ is, the solution $\hat{\x}$ produced by the $\ell_1$ minimization
satisfies
\begin{equation*}
 \|\x_K\|_1-\|\hat{\x}_K\|_{1} \leq
\frac{2}{C-1} \|\x_{\overline{K}}\|_1
\end{equation*}
 and
\begin{equation*}
 \|(\x-\hat{\x})_{\overline{K}}\|_1
\leq \frac{2C}{C-1} \|\x_{\overline{K}}\|_1,
\end{equation*}
if $\forall \w\in \mathbb{R}^n~\mbox{such that}~A\w=0$,
we have
 \begin{equation}
\|\x_K+\w_{K}\|_1+ \|\frac{\w_{\overline{K}}}{C}\|_1 \geq \|\x_K\|_1.
 \label{eq:Grasswthmeq1}
 \end{equation}
\end{thm}


 From \cite{D} and \cite{isitrobust}, if the matrix $A$ is sampled from an i.i.d. Gaussian ensemble, and $C=1$, for a single index set $K$, there exists a weak threshold $0<\mu_{W}<1$ such that if $\frac{|K|}{n} \leq \mu_{W}$, then with overwhelming probability as $n \rightarrow \infty$, the condition (\ref{eq:Grasswthmeq1}) holds for all $\w\in \mathbb{R}^n~\mbox{satisfying}~A\w=0$. Now if we take a single index set $K$ with cardinality $\frac{|K|}{n}=(1-\varpi){\mu_{W}}$, we would like to derive a characterization of $C$, as a function of $\frac{|K|}{n}=(1-\varpi){\mu_{W}}$, such that the condition (\ref{eq:Grasswthmeq1}) holds for all $\w\in \mathbb{R}^n~\mbox{satisfying}~A\w=0$. This is stated in the following theorem.

\begin{thm} \cite{XuScaingLaw}
Assume the $m\times n$ measurement matrix $A$ is sampled from an i.i.d. Gaussian ensemble, and let $K$ be a single index set with $\frac{|K|}{n}=(1-\varpi){\mu_{W}}$, where ${\mu_{W}}$ is the weak threshold for ideally sparse signals and $\varpi$ is any real number between $0$ and $1$. We also let $\x$ be an $n$-dimensional signal vector with $\x_{K}$ being an arbitrary but fixed signal component. Then with overwhelming probability, the condition (\ref{eq:Grasswthmeq1}) holds for all $\w \in \mathbb{R}^n~\mbox{satisfying}~A\w=0$, under the parameter $C=\frac{1}{\sqrt{1-\varpi}}$.
\label{thm:scalinglaw}
\end{thm}

\section{Non-isometric Sensing Matrices}
\label{sec:matrices}
It is well known that compressed sensing matrices $A$ should be isometric. For example, restricted isometry condition is a widely used condition \cite{CT} to prove that $\ell_1$ minimization provides performance guarantees of successfully recovering sparse signals. However, in this paper, we propose to use non-isometric matrices for compressed sensing.

As is often the case in compressed sensing \cite{CT,D}, sensing matrices consist of i.i.d. elements following a certain distribution, for example, the Gaussian distribution. For such a matrix $A$, with high probability, different columns will be roughly equal to each other in length. In our design, we adopt a weighted version of traditional sensing matrices by multiplying each column of $A$ with a randomly generated number. Let $A$ be an usual compressed sensing matrix, and let $AW$ denote our proposed sensing matrix. Then
$$\A W=\A \times  W, $$
where $W$ is a diagonal matrix with each diagonal element being nonzero. In this paper, we generate each of the diagonal elements from the standard Gaussian distribution $\mathcal{N}(0,1)$.Then the measurement results are given by
$$\y=\A W \x^{true},$$
where $\x^{true}$ is the original sparse signal.
Note that if we denote $\x=W\x^{true}$, then we have
$$\y=\A\x.$$

\section{Modified Reweighted $\ell_1$ Algorithm}
\label{sec:Algorithm}
Our proposed algorithm is a modified iterative reweighted algorithm tailored to non-isometric matrices. To find the sparse signal, instead of solving the following $\ell_0$ minimization problem
\begin{equation}
\min{ \|\z\|_0}~~\text{subject to}~~ \A W\z
= \y,
\end{equation}
we solve a modified but equivalent problem
\begin{equation}
 \min{ \|W\z\|_0}~~\text{subject to}~~ \A W\z= \y,
\label{eq:modifiedL0}
\end{equation}
because $W\z$ has the same support as $\z$.

Replacing $W\z$ with $\z$, (\ref{eq:modifiedL0}) further reduces to
\begin{equation}
 \min{ \|\z\|_0}~~\text{subject to}~~ \A \z= \y.
\label{eq:simplestL0}
\end{equation}

On the surface, this is nothing but a usual sparse signal recovery problem, where $\A$ is used to generate a measurement vector $\A \z$. However, in our proposed scheme, simply because $\A W$ is used as the sensing matrix, nonzero elements of $W\x^{true}$ (appearing in $\y= \A (W\x^{true})$) can not be constant-modulus; moreover, nonzero elements of $W\x^{true}$ follow an amplitude probability density function $f(\cdot)$ with $f(0)>0$.

This inspires us to propose the following reweighted $\ell_1$ minimization algorithm, modified from the reweighted algorithms from \cite{Icassp reweighted l_1, Isit reweighted l_1}. The algorithm consists of two $\ell_1$ minimization steps: a standard one and a weighted one. The input to the algorithm is the vector $\y=\A W\x^{true}$, where $\x^{true}$ is a $k$-sparse
signal with $k=(1+\epsilon_0)\mu_W(\delta)n$, and the output is an
approximation $\x^*$ to the unknown vector $\x$. We assume that $k$,
or an upper bound on it, is known. We remark that this is not a critical assumption, because there are at most $n$ possibilities for the sparsity. Also $\omega>1$ is a
predetermined weight.
%
\begin{algorithm}
\text{}
\begin{enumerate}
\item Solve the $\ell_1$ minimization problem:
\begin{equation}
\hat{\x} = \arg{ \min{ \|\z\|_1}}~~\text{subject to}~~ \A\z
= \y.
\end{equation}
\item Obtain an approximation for the support set of $\x$:
find the index set $L \subset \{1,2, ..., n\}$ which corresponds to
the largest $k$ elements of
$\hat{\x}$ in magnitude.
\item Solve the following weighted $\ell_1$ minimization problem and declare the solution as output:
 \beq
\tilde{\x} = \arg{\min\|\z_L\|_1+\omega\|\z_{\overline{L}}\|_1}~~\text{subject to}~~ \A\z
= \y.
\label{eq:weighted l_1}
\eeq

\item $\x^{*}=W^{-1}\tilde{\x}.$
\end{enumerate}
\label{alg:modmain}
\end{algorithm}

The idea behind the algorithm is as follows. In the first step
we perform a standard $\ell_1$ minimization. If the sparsity of the
signal is beyond the weak threshold $\mu_W(\delta)n$, then $\ell_1$ minimization is not capable of recovering the signal. However, we can use its output to identify an index set $L$ in which most elements correspond to the nonzero elements of $\x$. We
finally perform a weighted $\ell_1$ minimization by penalizing those
entries of $\x$ that are not in $L$ because they have a
lower chance of being nonzero elements.

In the next sections we formally prove that, for certain classes of signals, Algorithm
\ref{alg:modmain} has a recovery threshold beyond that of standard
$\ell_1$ minimization, even for sparse signals with constant-modulus nonzero elements.
By denoting $\x=W\x^{true}$ and recognizing $\y=A\x$, the phase transition improvement results of \cite{XuScaingLaw} apply to $\x$, since $\x$ now has an amplitude probability density function $f(\cdot)$ such that $f(0)\neq 0$. Once we can recover $\x$, $\x^{true}=W^{-1} \x$ will also be successfully recovered. 
For readers' convenience, we outline the reasoning steps of \cite{XuScaingLaw} and \cite{iterativereweightedjournal} in Sections \ref{sec:robustness} and \ref{sec:perfect recovery}. In Section
\ref{sec:robustness}, we prove that there is a large overlap between
the index set $L$, found in Step 2 of the algorithm, and the support set
of the unknown signal $\x$ (denoted by $K$)---see Theorem \ref{thm:l_1
  support recovery}. Then in Section
\ref{sec:perfect recovery}, we show that the large overlap between $K$
and $L$ can result in perfect recovery of $\x$, beyond the standard
weak threshold, when a weighted $\ell_1$ minimization is used in Step
3.


\section{Approximate Support Recovery, Steps 1 and 2 of the Algorithm \cite{XuScaingLaw}}
\label{sec:robustness}

In this section, we carefully study the first two steps of Algorithm
\ref{alg:modmain}. The unknown signal $\x$ is assumed to be a $k$-sparse vector with support set $K$, where
$k=|K|=(1+\epsilon_0)\mu_{W}(\delta)n$, for some $\epsilon_0>0$. 
The set $L$, as defined in the algorithm, is in fact the
$k$-support set of $\hat{\x}$. We show that for small enough
$\epsilon_0$, the intersection of  $L$ and $K$ is very large with high
probability, so that $L$ can be counted as a good
approximation to $K$. The main results are summarized in Theorem  \ref{thm:l_1 support recovery} \cite{XuScaingLaw}.
\begin{thm}\cite{XuScaingLaw} [Support Recovery]
Let $\A$ be an i.i.d. Gaussian $m\times n$ measurement matrix with
$\frac{m}{n}=\delta$. Let $k=(1+\epsilon_0)\mu_{W}(\delta)$ and $\x$
be an $n\times1$ random $k$-sparse vector whose nonzero element amplitude follows the distribution of $f(x)$. Suppose that
$\hat{\x}$ is the approximation to $\x$ given by the $\ell_1$ minimization, namely $\hat{\x}=argmin_{\A\z=\A\x}\|\z\|_1$. Then, for any $\epsilon_{0}>0$ and for all $\epsilon>0$, as $n\rightarrow\infty$,
\beq
\small
\Prob(\frac{|supp(\x) \cap supp_k(\hat{\x})|}{k} -(1-F(y^*))
>-\epsilon)\rightarrow 1,
\label{eq:support recovery}
\eeq
where $y^*$ is the solution to $y$ in the equation $\int_{0}^{y} xf(x) dx=\zeta(\epsilon_0)$.

Moreover, if the integer $t\geq 0$ is the smallest integer for which the amplitude distribution $f(x)$ has a nonzero $t$-th order derive at the origin, namely $f^{(t)}(0) \neq 0$, then as $\epsilon_{0} \rightarrow 0$, with high probability,
\beq
\small
\frac{|supp(\x) \cap supp_k(\hat{\x})|}{k}= 1-O(\epsilon_{0}^{\frac{1}{t+2}}).
\eeq
\label{thm:l_1 support recovery}
\end{thm}

The proof of Theorem \ref{thm:l_1 support recovery} relies on the scaling law for recovery stability. Note that if $\epsilon_0\rightarrow0$, then  Theorem \ref{thm:l_1 support recovery} implies that $\frac{|K\cap L|}{k}$ becomes arbitrarily close to 1. We can also see that the support recovery is better when the probability distribution function of $f(x)$ has a lower order of nonzero derivative.  This is consistent with the better recovery performance observed for such distributions in simulations of the iterative reweighted $\ell_{1}$ minimization algorithms \cite{CWB07}.

 \section{Perfect Recovery, Step 3 of the Algorithm \cite{XuScaingLaw}}
 \label{sec:perfect recovery}
 In Section \ref{sec:robustness} we showed that. if
 $\epsilon_0$ is small, the $k$-support of $\hat{\x}$, namely
 $L=supp_k(\hat{\x})$, has a significant overlap with the true support of
 $\x$. The scaling law gives a quantitative lower bound on the size of this overlap
 in Theorem \ref{thm:l_1 support recovery}. In Step 3 of
 Algorithm \ref{alg:modmain}, weighted $\ell_1$ minimization is used,
 where the entries in  $\overline{L}$ are assigned a higher weight
 than those in $L$. In \cite{isitweighted}, we have been able to
 analyze the performance of such weighted $\ell_1$ minimization
 algorithms. The idea is that if a sparse vector $\x$ can
 be partitioned into two sets $L$ and $\overline{L}$, where in one set
 the fraction of non-zeros is much larger than in the other set, then
 (\ref{eq:weighted l_1}) can increase the recovery threshold of $\ell_{1}$ minimization.
\begin{thm} \cite{isitweighted}
Let $L\subset \{1,2,\cdots,n\}$ , $\omega>1$ and the fractions
$f_1,f_2\in[0,1]$ be given. Let $\gamma_1 = \frac{|L|}{n}$ and
$\gamma_2=1-\gamma_1$. Measurement matrices $\A$ have i.i.d. $\mathcal{N}(0,1)$ Gaussian elements. There exists a threshold
$\delta_c(\gamma_1,\gamma_2,f_1,f_2,\omega)$ such that, with overwhelming
probability, a sparse vector $\x$ with \emph{at least}
$f_1\gamma_1n$ nonzero entries over the set $L$, and \emph{at most}
$f_2\gamma_2n$ nonzero entries over the set $\overline{L}$ can be
perfectly recovered using
$\min_{\A\z=\A\x}\|\z_L\|_1+\omega\|\z_{\overline{L}}\|_1$, where $\A$
is a $\delta_cn\times n$ matrix with i.i.d. Gaussian entries. Furthermore, for appropriate $\omega$,
\[
\mu_W(\delta_c(\gamma_1,\gamma_2,f_1,f_2,\omega))<f_1\gamma_1+f_2\gamma_2,
\]
i.e., standard $\ell_1$ minimization using a $\delta_cn\times n$
measurement matrix with i.i.d. Gaussian entries cannot recover such $x$.
\label{thm:delta}
\end{thm}

~\\
To apply Theorem \ref{thm:delta} to the approximate support recovery property, we should consider all the possibilities for $supp_k(\hat{\x})$. In fact, there are at most $\binom{k}{|supp(\x) \cap supp_k(\hat{\x})|}\binom{n-k}{k-|supp(\x) \cap supp_k(\hat{\x})|}$ possibilities for $supp_k(\hat{\x})$. When \beq
\small
\frac{|supp(\x) \cap supp_k(\hat{\x})|}{k}= 1-O(\epsilon_{0}^{\frac{1}{t+2}}),
\eeq
 a union bound over all the possibilities will be overwhelmed by the negative exponent of the failure probability in Theorem \ref{thm:delta} as $\epsilon_0 \rightarrow 0$, thus leading to Theorem \ref{thm: final thm}.

The main threshold improvement result is summarized in the following theorem \cite{XuScaingLaw}. For a detailed proof of this theorem, the readers can refer to \cite{iterativereweightedjournal}.

\begin{thm}[Perfect Recovery]
Let $\A$ be an $m\times n$ i.i.d. Gaussian matrix with
$\frac{m}{n}=\delta$. If $\delta_c(\mu_{W}(\delta),1-\mu_{W}(\delta),1,0,\omega) < \delta$,
then there exist $\epsilon_0>0$ and $\omega>0$ such that, with high probability as $n$ grows to infinity, Algorithm
\ref{alg:modmain} perfectly recovers a random
$(1+\epsilon_0)\mu_{W}(\delta)n$-sparse vector with i.i.d. nonzero
entries following an amplitude distribution whose pdf has a nonzero derive of some finite order at the origin.
\label{thm: final thm}
\end{thm}

\section{Simulation Results}
\label{sec:simulation}
In this section, we present simulation results of the phase transition performance of our new algorithm. In our simulation, we consider sparse signals with constant-modulus nonzero elements, for which conventional iterative reweighted algorithms failed to elevate the phase transition performance \cite{CWB07}. We also consider sparse signals with Gaussian nonzero elements. Our simulation results indeed show that the new algorithm indeed universally elevates the phase transition performance of compressed sensing, no matter what amplitude distribution the nonzero elements follow.

In the first simulation, the signal vector dimension $n$ is chosen to be $1000$, and the number of measurements $m=500$. The nonzero elements of sparse signals take value $+1$ or $-1$ independently with equal probability. We remark, however, that the decoder does not know the magnitude of the constant-modulus nonzero elements or whether the nonzero elements are constant-modulus. For sparse signals with constant-modulus nonzero elements, it was noted \cite{CWB07} that iterative reweighted $\ell_1$ minimization algorithms have almost the same phase transition performance as plain $\ell_1$ minimization, and so we only simulate plain $\ell_1$ minimization algorithms for comparison with our new algorithm.

  For one simulated curve, we use measurement matrices $\A$ with i.i.d. zero-mean Gaussian elements $\mathcal{N}(0,1)$, and plain $\ell_1$ minimization was used to recover the sparse signals. For another simulated curve, non-isometric matrices $AW$ are generated by multiplying columns of $A$ with independent  $\mathcal{N}(0,1)$ Gaussian random variables, and our proposed new algorithm is used to recover the sparse signals. For both curves, we simulate $100$ random examples for each sparsity level, and the decoding is declared successful if the decoding error $\|\x^{true}-{\x}^{*}\|^2 \leq 10^{-6}$.

\begin{figure}[t]

\centering
\includegraphics[width=3.75in, height=3in]{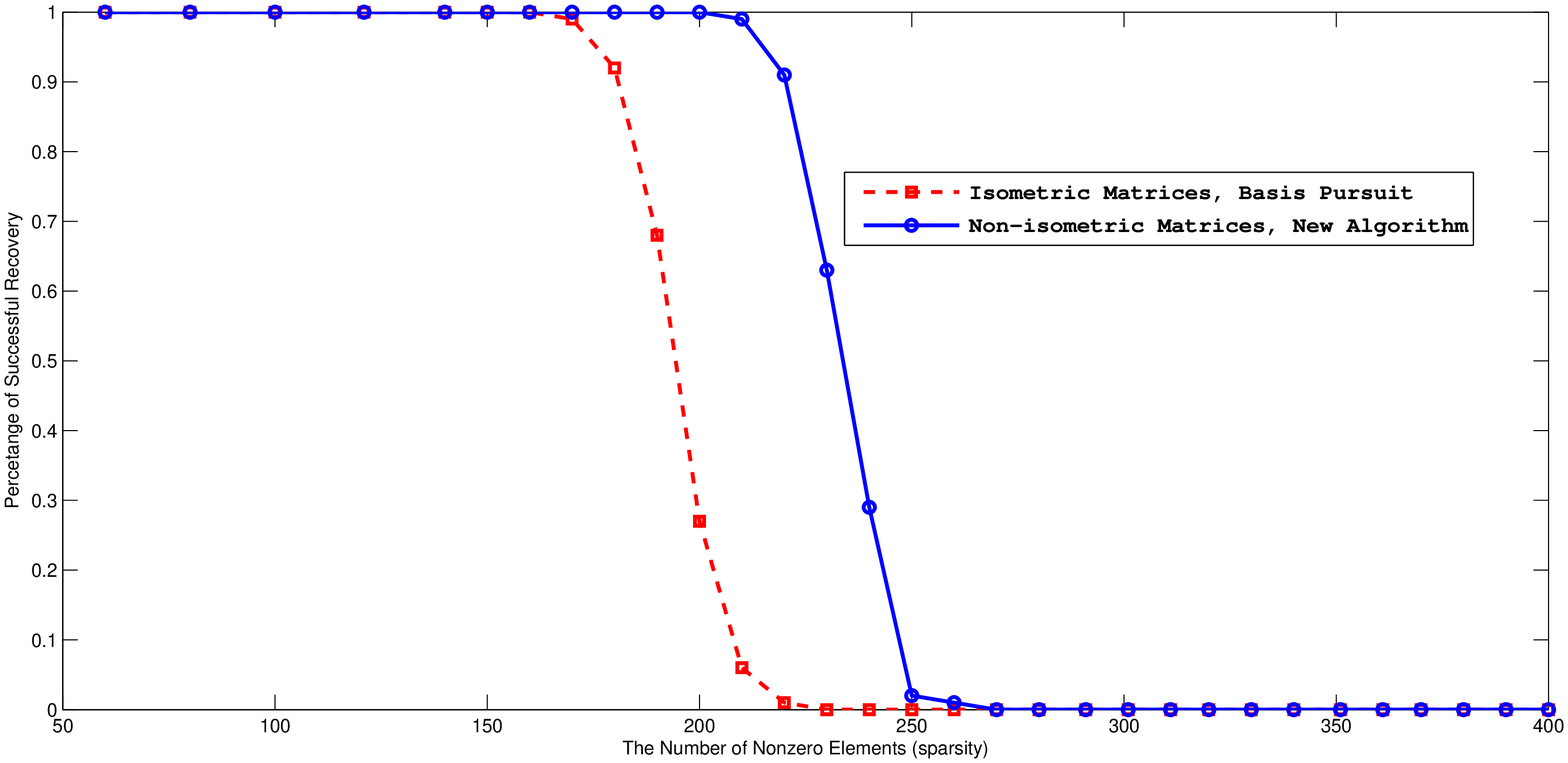}
\caption{Percentage of Successful Recovery versus sparsity, for sparse signals with constant-modulus non-zero elements, $n=1000$ and $m=500$}\label{fig:successraten1000m500}
\end{figure}

Figure \ref{fig:successraten1000m500} shows that the phase transition threshold for plain $\ell_1$ minimization is around $\frac{k}{n}=0.17$. For non-isometric matrices, and our new algorithm, the threshold is around $\frac{k}{n}=0.21$, a $23\%$ increase over plain $\ell_1$ minimization.

\begin{figure}[t]
\centering
\includegraphics[width=3.75in, height=3in]{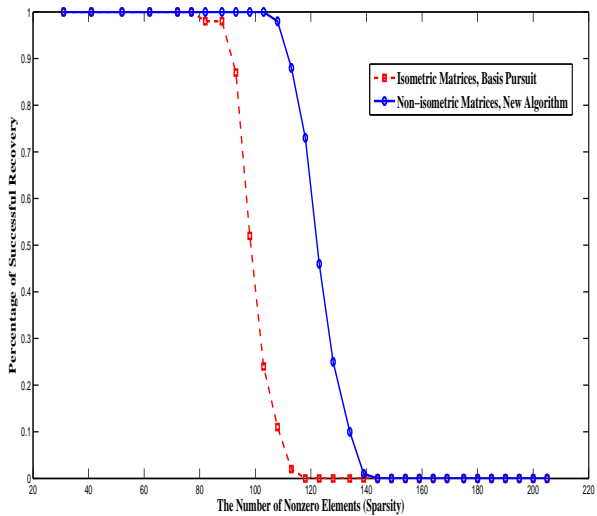}
\caption{Percentage of Successful Recovery versus sparsity, for sparse signals with Gaussian nonzero elements.  $n=512$ and $m=256$.}\label{fig:successraten512m256}
\end{figure}

In the second the simulation, we adopt the same setting as the first simulation, except that $n=512$, $m=256$ and the nonzero elements of sparse signals are taken as i.i.d. $\mathcal{N}(0,1)$ Gaussian random variables. Figure \ref{fig:successraten512m256} shows that the phase transition threshold for plain $\ell_1$ minimization is around $\frac{k}{n}=0.17$. For non-isometric matrices, and our new algorithm, the threshold is around $\frac{k}{n}=0.21$, also a $23\%$ increase over plain $\ell_1$ minimization.

\section{Acknowledgement}
 We thank E. Candes, A. Khajehnejad, Babak Hassibi and S. Avestimehr for helpful discussions.

\bibliographystyle{IEEEbib}

\end{document}